\documentstyle[aps,epsfig]{revtex} 
\begin{document} 
\preprint{} 
\draft 

\title{Continuous quantum measurement of a Bose-Einstein condensate:\\
                a stochastic Gross-Pitaevskii equation              } 
\author{ Diego A. R. Dalvit${}^{1}$, 
         Jacek Dziarmaga$^{1,2}$, 
         and Roberto Onofrio$^{1,3}$ } 
\address{ 
${}^1$ Los Alamos National Laboratory, Los Alamos, 
       New Mexico 87545 \\ 
${}^2$ Intytut Fizyki Uniwersytetu Jagiello\'nskiego, 
       ul. Reymonta 4, 30-059 Krak\'ow, Poland \\ 
${}^3$ Dipartimento di Fisica "G. Galilei", 
       Universit\`a di Padova, Padova 35131, Italy \\
       and 
       INFM, Sezione di Roma "La Sapienza", Roma 00185, Italy} 
\date{December 12, 2001}
\maketitle

\begin{abstract}
We analyze the dynamics of a Bose-Einstein condensate undergoing a
continuous dispersive imaging by using a Lindblad operator formalism.
Continuous strong measurements
drive the condensate out of the coherent state description assumed within
the Gross-Pitaevskii mean-field approach. Continuous weak measurements
allow instead to replace, for timescales short enough, the exact problem
with its mean-field approximation through a stochastic analogue of the
Gross-Pitaevskii equation. The latter is used to show the unwinding of a
dark soliton undergoing a continuous imaging.
\end{abstract}


\section{Introduction}

  The interplay between quantum and classical descriptions of the physical
world and the role of the measurement process are still at the heart of
the understanding of quantum mechanics \cite{Wheeler}. The related
theoretical debate has been greatly enriched in the last decades by the
realization of new experimental techniques aimed at producing quantum
states without classical analogue, such as entangled or squeezed states,
or to explore phenomena which are intrinsically quantum mechanical, like
quantum jumps.  All this occurred also having in mind practical
implications, such as the improvement of sensitivity of various devices
operating at or near the quantum limit \cite{Braginsky,Santarelli}.

  Recently, the production of a novel state of matter - Bose-Einstein
condensates of dilute atomic gases - has opened a new road to explore
macroscopic quantum phenomena with the precision characteristic of atomic
physics \cite{Inguscio}. Bose-Einstein condensates are naturally produced
by cooling down atomic gases at ultralow temperature with the phase
transition occurring in the 100 nK - 1$\mu$K temperature range, 
{\it i.e.} when the
thermal de Broglie wavelength becomes comparable to the average spacing
between the atoms of a dilute (peak density $10^{13} - 10^{15}$
atoms/cm${}^3$) trapped gas.  Usually samples of Bose-Einstein condensates
are made by $10^3-10^8$ atoms in the case of alkali species
\cite{Inguscio}, and $10^9$ or more in the case of hydrogen
\cite{Kleppner}.  Their intrinsically small heat capacity does not allow
for a direct manipulation and probing with material samples, such as
microtips or nanostructures, since the thermal contact with the latter
will induce a sudden vaporization of the Bose sample. Thus, manipulation
and probing of Bose-Einstein condensates has been achieved so far only by
using light beams.  These probes can be classified according to the
resonant or nonresonant nature of their interaction with the measured
atomic sample.  In the former case, the condensate interacts with a laser
beam resonant (or close to resonance) with a selected atomic transition.  
The output beam is attenuated proportionally to the optical thickness of
the condensate (also called column density), {\it i.e.} the condensate
density integrated along the line of sight of the impinging beam.  The
absorption of the photons leads to recoil of the atoms then strongly
perturbing the condensate. For typical values of intensity and duration of
the probe light the condensate is strongly heated and a new replica has to
be produced to further study its dynamics. From the viewpoint of quantum
measurement theory this measurement is of type-II since it destroys the
state of the observed system and forbids the study of the dynamics of a
single quantum system \cite{Pauli}. An alternative technique, called
dispersive imaging, allows for repeated measurements on a Bose-Einstein
condensate and, as an extreme case, its continuous monitoring.  In this
measurement scheme off-resonance light is scattered by the condensate,
thereby locally inducing optical phase-shifts which can be converted into
light intensity modulations by homodyne or heterodyne techniques, for
instance by using phase-contrast \cite{Andrews} or interference techniques
\cite{Kadlecek}.  Since the laser beam is off-resonant the absorption rate
is small and the heating of the condensate is accordingly small. Thus,
multiple shots of the same condensate can be taken - a type-I measurement
in the quantum measurement theory language - allowing the study of the
dynamics on the same sample. This has allowed to overcome the unavoidable
shot-to-shot fluctuations always present in the production of different
samples of Bose-Einstein condensates. Several phenomena whose observation
rely on imaging at high accuracy the condensate, such as its formation in
non-adiabatic conditions \cite{Miesner}, short \cite{Andrews1} and long
\cite{Stamper} wavelength collective excitations, superfluid dynamics
\cite{Onofrio}, vortices formation and decay \cite{Matthews,Abo}, can be
successfully studied with this technique.

  The repeated nondestructive monitoring lends itself to another question:  
{\sl is the measurement process influencing the dynamics of the
condensate?} Answers to this question are important also for disentangling
the intrinsic condensate dynamics from the artifacts induced by the
underlying measurement process. As we will see in this paper, the effect
of the measurement can also be intentionally amplified to allow for
unusual manipulation of the condensate itself. Dispersive imaging, in its
more idealized form, is preserving the number of atoms and therefore
represent a particular example of quantum nondemolition (QND) measurement
\cite{Braginsky,Caves,Bocko}.  We know that even quantum nondemolition
measurements affect the state of the observed system, their unique feature
being that the nondemolitive observable maintains the same {\it average}
value, albeit the probability distribution of its outcome can be affected
as well as the average values of all the conjugate observables.  Thus, we
do expect that during the nondemolitive measurement of the condensate atom
number there will be a measurement-induced nontrivial dynamics for both
the variance of the monitored observable and the average values of all
conjugate observables.  Besides gaining insight into the dynamics of the
measurement process, our model allows for the description of the {\it
irreversible} driving of the condensate toward nonclassical states.

In this paper we try to answer the question formulated above by
building a realistic 
model for dispersive measurements of the condensate, extending
the results reported in \cite{Dalvit} to the weak measurement regime.  
The plan of the paper is as follows. In Section II we introduce the
dispersive coupling between atoms and light and derive the reduced master
equation for the Bose condensate tracing out the variables of the
electromagnetic field degrees of freedom. Under controllable
approximations we obtain a Lindblad equation which allows to estabilish
the rates for phase diffusion and depletion of the condensate during the
dispersive imaging.  In Section III we unravel the
Lindblad equation in terms of a stochastic differential equation that, in
the unmeasured case, corresponds to the description of a single N-body
wavefunction.  A solution of the stochastic N-body equation is discussed
in the limit of strong continuous measurement, leading to the squeezing of
number fluctuations, the main result described in
\cite{Dalvit}. In the opposite limit of weak measurement and for an
initial mean-field state this stochastic equation becomes the stochastic
counterpart of the Gross-Pitaevskii equation, as discussed in Section IV.
Its limit of validity is discussed in
Section V by comparing its evolution for various parameters versus the
exact evolution in the simple situation of a two mode system 
schematizing a
condensate in a double well potential. This allows us to analyze, in
Section VI, the effect of the measurement on the evolution of a condensate
initially prepared in a soliton state. More general considerations on the
potentiality of such an approach and its consequences are finally outlined
in the conclusions.

\section{Master equation for dispersive imaging of a Bose condensate}

Our main goal in this Section is to include the atom-photon interaction 
present in the dispersive imaging of a Bose-Einstein condensate into its 
intrinsic dynamics. First attempts in this direction have been discussed in 
the prototypical situation of a two mode condensate in \cite{Walls,Milburn}. 
Let us start the analysis with the effective interaction Hamiltonian between 
the off-resonant photons and the atoms, written as
\begin{equation} 
H_{\rm int}= {{\epsilon_0 \chi_0} \over 2} 
\int d^3x~ n({\bf x}) : {\bf E}^2 : ,  
\label{Intham} 
\end{equation} 
where $n$ is the density operator of the atomic vapor, and 
${\bf E}$ the electric field due to the intensity $I$ of the
incoming light.  The coefficient $\chi_0$ represents the
effective electric susceptibility of the atoms defined as
$\chi_0=\lambda^3 \delta/2 \pi^2 (1+\delta^2)$, where we
have introduced the light wavelength $\lambda$ and the
light detuning measured in half-linewidths $\Gamma/2$ of
the atomic transition, $\delta=(\omega-\omega_{\rm at})/(\Gamma/2)$. 

We express the electric field in terms of creation and annihilation
operators. In the Coulomb gauge it takes the form
\begin{equation}
{\bf E}({\bf x},t) = i \sum_{{\bf k}} 
\sqrt{\frac{\hbar \omega_{\bf k}}{2 \epsilon_0 L^3}}
( a_{\bf k} e^{-i c k t + i {\bf k} {\bf x} } -
  a_{\bf k}^{\dagger} e^{i c k t - i {\bf k} {\bf x}}) ,
\end{equation}
where $\omega_{\bf k} = c |{\bf k}|$, $[a_{\bf k},a_{\bf k'}^{\dagger}] = 
\delta_{{\bf k},{\bf k'}}$, and $L^3$ is the quantization volume.

Equation (\ref{Intham}) allows us to write the reduced master equation for 
the atomic degrees of freedom by a standard technique, {\it i.e.} by 
tracing out the photon degrees of freedom \cite{Carmichael}. 
The decoupling between the two relevant timescales for the photons (settled 
by the the lifetime of spontaneous emission, of order of tens ns) and for the 
atoms (related to the oscillation period in the trapping potential, of the 
order of ms), allows to use the Born-Markov approximation.
Thus we get the master equation for the reduced density matrix $\rho$ 
of the condensate that, in the interaction picture, is written as 

\begin{equation}
\frac{d \rho}{dt} = \frac{i}{\hbar} {\rm Tr}_{\rm R} 
[ \rho(t) \otimes \rho_{\rm R}(t), H_{\rm int}] - 
\frac{1}{\hbar^2} {\rm Tr}_{\rm R}
[H_{\rm int} (t), [ \int_{-\infty}^t dt' H_{\rm int}(t'), 
\rho(t) \otimes \rho_{\rm R}(t) ]] .
\label{ME}
\end{equation}
The last term on the righthandside contains two different contributions 
both of Lindblad type, $L_1 \rho $ and $L_2 \rho$. As we will see soon, 
the former preserves the number of atoms in the condensate and is responsible 
for phase diffusion phenomena, while the latter changes the number of atoms 
leading to its depletion. 

Let us first concentrate on phase-diffusion which is a number conserving
mechanism. To calculate it we insert the interaction Hamiltonian into the 
last term of Eq.(\ref{ME}), with $n$ the condensate density operator. 
By introducing the Fourier transform of the density operator such that

\begin{equation}
n({\bf x}) = \sum_{\bf q} e^{i {\bf q}{\bf x}} 
\frac{\tilde{n}({\bf q})}{\sqrt{L^3}} ,
\end{equation}
we obtain
\begin{eqnarray}
L_1 \rho &=& \frac{\pi \chi_0^2}{2 L^3 c} 
\sum_{{\bf k}, {\bf p}} \sqrt{\omega_{{\bf k}} \omega_{{\bf p}}} 
\delta(k-p) \sum_{{\bf k'}, {\bf p'}} \sqrt{\omega_{{\bf k'}} 
\omega_{{\bf p'}}} e^{i c t (k'-p')} \times \nonumber \\
&&
\left[
\tilde{n}({\bf k'}-{\bf p'}) \tilde{n}({\bf k}-{\bf p}) \rho
\langle a_{\bf k'}^{\dagger} a_{\bf p'} a_{\bf k}^{\dagger} 
a_{\bf p} \rangle -\tilde{n}({\bf k'}-{\bf p'}) \rho \tilde{n}
({\bf k}-{\bf p}) \langle a_{\bf k}^{\dagger} a_{\bf p} 
a_{\bf k'}^{\dagger} a_{\bf p'} \rangle
 \nonumber \right. \\
&& \left.
- \tilde{n}({\bf k}-{\bf p}) \rho \tilde{n}({\bf k'}-{\bf p'}) 
\langle a_{\bf k'}^{\dagger} a_{\bf p'} a_{\bf k}^{\dagger} a_{\bf p} 
\rangle
+
\rho \tilde{n}({\bf k}-{\bf p}) \tilde{n}({\bf k'}-{\bf p'}) 
\langle a_{\bf k}^{\dagger} a_{\bf p} a_{\bf k'}^{\dagger} 
a_{\bf p'} \rangle
\right] ,
\end{eqnarray}
where $\langle \dots \rangle = {\rm Tr}_{\rm R}[\rho_{\rm R} \ldots ]$.
The photons are assumed to be in a coherent plane wave state with
momentum along the impinging direction, corresponding to a
wavevector ${\bf k}=k_0 \hat{z}$ orthogonal to the imaging
plane $x-y$.  Hence

\begin{equation}
\langle a_{\bf k'}^{\dagger} a_{\bf p} \rangle = \frac{L^3 I} {c^2 \hbar k_0}
\delta_{{\bf k'},{\bf p}} \delta_{{\bf p},{\bf k}_0},
\end{equation}
where we have written the mean numbers of photons in mode $k_0$ in terms
of the intensity of the incoming beam. Expressing the expectation values
in normal ordering and using the fact that phase-diffusion processes conserve
the number of bosons in the condensate, it is possible to show that all normal
ordered expectation values involving four operators cancel exactly, obtaining

\begin{equation}
L_1 \rho = \frac{ \pi \chi_0^2 I k_0 }{2 \hbar c}
\left( \frac{L}{2 \pi} \right)^3
\int d^3k\; \delta(| {\bf k} + {\bf k_0}| - k_0 )\;
[\tilde{n}(- {\bf k}), [ \tilde{n}({\bf k}), \rho]],
\end{equation}
where we have used the continuum limit $\sum_{\bf k} \rightarrow (L/2
\pi)^3 \int d^3k$. Unless tomographic techniques are used as for instance
in \cite{Andrews2}, the image results from a projection of the condensate
onto the $x-y$ plane, by integrating along the $z$ direction. This demands
to project the dynamics of the condensate into the imaging plane. In order
to write a closed 2D master equation to describe the $x-y$ dynamics we
assume the condensate wavefunction to be factorizable as
$\psi(x,y,z)=\phi(x,y) \Lambda(z)$.  Such factorization holds if the
confinement in the $z$-direction is strong enough to make the
corresponding mean-field energy negligible with respect to the energy
quanta of the confinement, {\it i.e.} $\hbar \omega_z >> g \tilde{\rho}$, where
$g=4 \pi \hbar^2 a/m$, with $a$ the s-wave scattering length, 
$\tilde{\rho}$ the
condensate density, and $\omega_z$ the angular frequency of the
confinement harmonic potential along the $z$ direction, as recently
experimentally demonstrated in \cite{Gorlitz}.  We write $\tilde{n}({\bf
k}) = \tilde{n}({\bf k}_{\perp}) \tilde{n}(k_z)$ and we will use a
Gaussian ansatz for the density profile along the $z$ direction, namely
\begin{equation} \tilde{n}(k_z) = \sqrt{\frac{2 \pi}{L}} e^{- \frac{\xi^2
k_z^2}{2}} , \end{equation} where $\xi$ is the lengthscale of the
condensate in the $z$ direction, the width of the Gaussian state
$\Lambda(z)$ under the abovementioned approximation.  The effective 2D
nonlinear coupling strength is $g_{2D}=g \int dz |\Lambda(z)|^4 =g
\sqrt{\pi}/\xi$. Consequently

\begin{equation}
L_1 \rho =
\frac{\pi \chi_0^2 I k_0}{2 \hbar c} 
\left( \frac{L}{2 \pi} \right)^2
\int d^2 k_{\perp}\; [ \tilde{n}(-{\bf k}_{\perp}), 
[\tilde{n}({\bf k}_{\perp}), \rho]] 
\left(\frac{L}{2 \pi}\right) 
\int dk_z\; \tilde{n}^2(k_z) 
\delta[|{\bf k} + {\bf k}_0| - k_0] .
\end{equation}
By assuming that the typical length of the BEC in the $z$ direction is
much larger than the wavelength of the incoming laser, i.e., $\xi \gg
\lambda$, we can calculate the value of the last integral, and it is equal
to $\exp(- \xi^2 {\bf k}_{\perp}^4 / 4 k_0^2)$.  Our final result for the
phase-diffusion contribution to the reduced master equation in the imaging
plane is written as

\begin{equation}
L_1 \rho = \int d^2r_1 \int d^2r_2\; K({\bf r}_1-{\bf r}_2)\; 
[ n({\bf r}_1),[n({\bf r}_2),\rho]] , 
\label{phasediff} 
\end{equation} 
where $n({\bf r})=\Psi^{\dagger}\Psi({\bf r})$ is the 2D density 
operator, and $K$ is the measurement kernel 
\begin{equation}
K({\bf r})= {{\pi \chi_0^2 k_0 I} \over 2 \hbar c} 
\int d^2k\; \exp(-\xi^2 k^4/4 k_0^2 + i {\bf k} {\bf r}) .  
\label{KERNEL}
\end{equation} 
Eq. (\ref{phasediff}) preserves the total number of atoms, and corresponds
to a quantum nondemolition coupling between the atom and the optical
fields \cite{Walls,Milburn,Onofrio1,Li,Leonhardt}. If the measurement
kernel were a local one, $K({\bf r}_1-{\bf r}_2) \simeq \delta({\bf r}_1
-{\bf r}_2)$, Eq.(\ref{phasediff}) would reduce to a Lindblad equation for
the measurement of an infinite number of densities $n({\bf r})$. This
assumes that no spatial correlation is established by the photon
detection. However, the ultimate resolution limit in the imaging system
depends on the photon wavelength, regardless of the pixel density of the
detecting camera. The resolution lengthscale follows from
Eq.(\ref{KERNEL}) as a width of the kernel 

\begin{equation} 
\Delta r=(2\pi^2\xi/k_0)^{1/2}=(\pi \xi \lambda)^{1/2}\;, 
\end{equation}
the geometrical average of the light wavelength and the condensate
thickness $\xi$. Eq. (\ref{phasediff}) can then be rewritten as $L_1 \rho =
\gamma_1 [n,[n,\rho]]$, where $\gamma_1$ is the phase diffusion rate,
given by

\begin{equation}
\gamma_1 = \frac{\pi \chi_0^2 k_0 I}{2 \hbar c} 
\int d^2k\;  e^{- \frac{\xi^2 k^4}{4 k_0^2}}
|\phi({\bf k})|^2  |\phi(-{\bf k})|^2 .
\end{equation}
We estimate its magnitude assuming a Gaussian profile in the x-y plane, i.e.,
$|\phi({\bf k})|^2 = e^{-\alpha^2 k^2/2}$, obtaining

\begin{equation}
\gamma_1 = \frac{ \pi^{5/2} \chi_0^2 k_0^2 I}{2 \hbar c \xi}
e^{\frac{\alpha^4 k_0^2}{\xi^2}} \left[ 1 - {\rm Erf} \left(
\frac{\alpha^2 k_0}{\xi} \right) \right]
\approx \frac{\pi^2 \chi_0^2 k_0 I}{2 \hbar c \alpha^2} ,
\label{gamma1}
\end{equation}
where the last step holds for a well localized condensate, {\it i.e.}
$\alpha ^2 k_0 / \xi \gg 1$. 

Let us now calculate the depletion
contribution to the master equation for the condensate. We split the field
annihilation operator into a term describing the condensate and another
associated to the non-condensed particles, {\it i.e.}, $\psi=\psi_{\rm C}
+ \psi_{\rm NC}$.  We shall assume that the non-condensed particles belong
to the continuum, so that their spectrum is the one of a free-particle
$\hbar \Omega_{\bf q} = \hbar^2 q^2/2 m$. Indeed, photons have large
momenta with respect to the momenta of trapped atoms, so even if a small
percentage of the photon momentum is absorbed by the atom, the atom is
promoted into a high energy, unbounded state.  Let us expand both field
operators in terms of annihilation operators as

\begin{eqnarray}
\psi_{\rm C}({\bf x},t) &=& b_{\rm C}  \phi_{\rm C}({\bf x}, t) 
\nonumber \\
\psi_{\rm NC}({\bf x},t) &=& \frac{1}{\sqrt{L^3}}
\sum_{\bf q} e^{-i \Omega_{\bf q} t + i {\bf q} {\bf x}}\; b_{\rm q} ,
\end{eqnarray}
where $\phi_{\rm C}$ is the condensate wave function. We get
\begin{eqnarray}
L_2 \rho &=& \frac{\pi \chi_0^2}{8 c L^6} 
\sum_{{\bf k}{\bf p}{\bf q}} \sqrt{\omega_{\bf k} \omega_{\bf p}} 
\sum_{{\bf k}'{\bf p}'{\bf q'}} \sqrt{\omega_{\bf k}' \omega_{\bf p}'}
\delta(\Omega_{{\bf q}'} - c k' + c p') \times \nonumber \\
&& \left\{
e^{i t (\Omega_{{\bf q}} - c k + cp )} \tilde{\phi}_{\rm C}(q+p-k)
\tilde{\phi}^*_{\rm C}(q'+p'-k')
{\rm Tr}_{\rm R} [ a^{\dagger}_{\bf p} a_{\bf k} b^{\dagger}_{\bf q} b_c , 
[a^{\dagger}_{{\bf k}'} a_{{\bf p}'} b_{{\bf q}'} b^{\dagger}_{\rm C}, \rho]]
\right. \nonumber \\
&& \left.
+
e^{-i t (\Omega_{{\bf q}} - c k + cp )} \tilde{\phi}^*_{\rm C}(q+p-k)
\tilde{\phi}_{\rm C}(q'+p'-k') 
{\rm Tr}_{\rm R}[ a^{\dagger}_{\bf k} a_{\bf p} b_{\bf q} b^{\dagger}_c , 
[a^{\dagger}_{{\bf p}'} a_{{\bf k}'} b^{\dagger}_{{\bf p}'} b_{\rm C}, \rho]]
\right\} .
\end{eqnarray} 
Here the trace is taken over the reservoir of the condensate, which in
this case consists of the non-condensed particles and the photons.  The
above expression contains depletion processes, in which a photon interacts
with a particle in the condensate and, as a result, that particle is
kicked out of the condensate. It also contains feeding processes, in which the
reverse mechanism may take place. When one assumes that in the initial state
of the non-condensate plane waves are empty, only the depletion process is
relevant. In this hypothesis we find
\begin{equation}
L_2 \rho = \gamma_2 \left(- b_{\rm C} \rho b^{\dagger}_{\rm C} + \frac{1}{2} 
\left\{ b^{\dagger}_{\rm C} b_{\rm C}, \rho \right\}
\right) ,
\end{equation}
where $\gamma_2$ is the depletion rate
\begin{equation}
\gamma_2 =  \frac{\pi \chi_0^2 I}{4 \hbar c L^3} 
\left( \frac{L}{2 \pi} \right)^6
\int d^3p d^3q \;
\omega_{\bf p} \;
| \tilde{\phi}_{\rm C}({\bf q}+{\bf p}-{\bf k}_0) |^2 \;
\delta(\Omega_{\bf q} - c k_0 + c p) .
\end{equation}
We evaluate $\gamma_2$ in the thick condensate limit ($\xi \gg k_0^{-1}$)
by approximating
$| \tilde{\phi}_{\rm C}({\bf q}+{\bf p}-{\bf k}_0)|^2 \approx
(2 \pi/L)^3 \delta({\bf q}+{\bf p}-{\bf k}_0)$, and then, by using the
fact that
$\Omega_{\bf q} \ll c k_0$, we obtain
\begin{equation}
\gamma_2 =  \frac{\chi_0^2 k_0^3 I}{8 \pi \hbar c} .
\label{gamma2}
\end{equation}
From Eqs.(\ref{gamma1},\ref{gamma2}) we see that the depletion rate is
much bigger than the phase diffusion rate $\gamma_2/\gamma_1 = \alpha^2
k_0^2/4 \pi^3 \gg 1$, in accordance with Ref.\cite{Leonhardt}. We estimate
the magnitude of both rates using the following parameters for the
condensate and its imaging, relevant for the case of $^{87}{\rm Rb}$:
$\lambda=780{\rm nm}$, $\chi_0=10^{-23}{\rm m}^3$, laser intensity
$I=10^{-4} {\rm mW}/{\rm cm}^2$, and a typical size in the x-y plane of
$\alpha = 10 \mu{\rm m}$. Then $\gamma_1 = 10^{-6} {\rm s}^{-1}$ and
$\gamma_2 = 10^{-5} {\rm s}^{-1}$, corresponding to phase diffusion and
depletion times of $t_1 = \gamma_1^{-1} = 10^5 {\rm s}$ and
$t_2=\gamma_2^{-1} = 10^4 {\rm s}$, respectively. Although the depletion
rate is larger that the phase diffusion rate, this last process can dominate
because of their different scaling with the total number of condensed
particles in the master equation. Indeed, the first is linear in $N$,
whereas the second one is quadratic, so for large number of particles
(such as $N=10^7$) phase diffusion occurs on a faster timescale than
depletion. For this reason we will focus in the
following on the phase diffusion contribution alone.


\section{Strong measurement: measurement-induced number squeezing}

We will consider in the following the effect of strong measurements on the
quantum state of the condensate. This has already been described in detail  
elsewhere \cite{Dalvit}, thus here we only summarize the main results 
and their link to the following considerations.
By neglecting the depletion term, the master equation for the condensate 
takes the form $\dot{\rho} = (-i/\hbar) [H,\rho] - L_1 \rho$. This equation
preserves the total number of atoms in the condensate and corresponds to a
quantum nondemolition measurement of the atomic density via the
optical fields. In order to get an insight into the equation, we introduce
a two-dimensional lattice with lattice constant set by the kernel
resolution $\Delta r$. In this way we get

\begin{equation}
\frac{d \rho}{dt} = - \frac{i}{\hbar} \left[
- \hbar \omega \sum_{\langle k,l \rangle} \Psi^{\dagger}_{k} 
\Psi_{k} + V, \rho \right] - S \sum_{l} [n_l,[n_l,\rho]] .
\label{lattice}
\end{equation}
Here $\Psi_l$ is an annihilation operator and $n_l = \Psi_l^{\dagger}
\Psi_l$ is the number operator at a lattice site $l$.  The frequency of
hopping between any nearest neighbor sites $\langle k,l \rangle$ is
$\omega\approx \hbar/ 2m \Delta r^2$, which is $\hbar^{-1}$ times the
characteristic kinetic energy.  The potential energy operator is $V=\sum_l
(U_l n_l+G n_l^2)$, where $U_l$ is the trapping potential and $G=g_{\rm
2D}/\Delta r^2$. The effective measurement strength is $S \approx \int
d^2r K({\bf r})/ \Delta r^2=(2 \pi/\Delta r)^2 (\pi \chi^2_0 k_0 I/2 \hbar
c)$. In order to solve this equation we use an unraveling in terms of pure
states $|\Psi\rangle$ such that $\rho = \overline{ |\Psi\rangle \langle
\Psi|}$, the overline denoting the average over the unraveling stochastic
realizations.  The pointer states of Eq. (\ref{lattice}) are not changed
by the chosen unraveling \cite{DDZ}. The pure states can be expanded in a
Fock basis per site, $|\Psi\rangle = \sum_{\{N_l\}} \psi_{\{N_l\}}
|\{N_l\} \rangle$ and the amplitudes $\psi_{\{N_l\}}$ satisfy the
following stochastic Schr\"odinger equation (written in the Stratonovich
convention)

\begin{equation}
\frac{d}{dt} \psi_{\{N_l\}} = -\frac{i}{\hbar}
\sum_{\{N'_l\}} h_{\{N_l,N'_l\}} \psi_{\{N'_l\}}
-\frac{i}{\hbar} V_{\{N_l\}} \psi_{\{N_l\}} 
+ \psi_{\{N_l\}} \sum_l \left[
-S(N_l-n_l)^2+ S\sigma_l^2+
(N_l-n_l)\theta_l \right] , 
\end{equation}
where the homodyne noises have averages $\overline{\theta_{l}(t)}=0$ and 
$\overline{\theta_{l_1}(t_1)\theta_{l_2}(t_2)}=
2S\delta_{l_1,l_2}\delta(t_1-t_2)$. Here
$n_l=\sum_{\{N_l\}}N_l|\psi_{\{N_l\}}|^2$, 
$\sigma_l^2=\sum_{\{N_l\}}(N_l-n_l)^2|\psi_{\{N_l\}}|^2$, 
$h$ is the matrix element of the hopping Hamiltonian and
$V_{\{N_l\}}=\sum_l(U_lN_l+GN_l^2)$ is the potential energy.  
 When there is no hopping term ($h=0$), an exact solution to this
equation as a product of Gaussian-like wavefunctions is written as
\begin{equation}
\psi_{\{N_l\}}(t)= e^{i\varphi_{\{N_l\}}}
e^{ -\frac{i}{\hbar}V_{\{N_l\}} t}
\prod_l \frac{ e^{-\frac{[N_l-n_l(t)]^2}{4\sigma_l^2(t)}} }
             {         [2\pi\sigma_l^2(t)]^{1/4}          } .
\end{equation}
The population mean value per site $n_l(t)$ does a random walk, while its
dispersion decreases as 
$\sigma_l^2(t)=\sigma_l^2(0)/[1+4\sigma^2_l(0)St]$. Here $\sigma_l^2(0)$
is the initial dispersion in the number of atoms per site, and it scales
as $N$ for an initial coherent state. Thus, the measurement drives the
quantum state of the condensate to a Fock state. When tunneling between
different lattice sites is allowed ($h \neq 0$), localization in a Fock
state is inhibited due to a competition between the measurement, which
drives localization, and hopping, which tries to drive the state of the
condensate towards coherent states. When measurement outweighs hopping,
the final state of the BEC is a number squeezed state. The timescale in
which squeezing is achieved is given by

\begin{equation}
t_{\rm SQ} = \frac{1}{n_l S} ,
\end{equation}
and the associated dispersion in the number of atoms per site for the
asymptotic squeezed state is

\begin{equation}
\sigma_l  = (\omega n_l/S)^{1/4} .
\end{equation}
In a coherent state the number fluctuations per site are Poissonian, 
$\sigma_l=n_l^{1/2}$. 
The state is squeezed when $(\omega n_l/S)^{1/4}<n_l^{1/2}$, 
i.e.  sub-Poissonian atomic number fluctuations.
This condition defines the strong measurement as

\begin{equation}
\frac{n_l S}{\omega} \; > \; 1 \;.
\end{equation} 
Number squeezing of a Bose condensate has been experimentally observed 
in an optical lattice in \cite{Orzel}. Our situation has an important
difference from the latter case:  since the squeezing is driven by the 
(Lindblad) measurement term the evolution into such states is irreversible, 
even after removal of the imaging photon field. From this viewpoint our 
squeezing technique is similar to the spin squeezing through quantum 
nondemolition measurements proposed in \cite{Kuzmich} and 
demonstrated in \cite{Kuzmich1}. Of course, the system will eventually 
drift towards coherent, classical states due to the interaction 
with the external environment and the related decoherence \cite{Zurek}, 
for instance due to the thermal component or residual background gas in the trapping volume.


\section{Weak measurement: stochastic Gross-Pitaevskii equation}

Unraveling the Lindblad equation derived above leads, in the 
limit of weak measurements, to a stochastic equation for the condensate 
wavefunction. In the mean-field approximation this equation becomes the 
analogue of the (deterministic) Gross-Pitaevskii equation for the unmeasured 
system. From Eqs.(\ref{phasediff},\ref{KERNEL}) and ignoring the depletion term
$-L_2\rho$ we obtain a continuum version of the master equation
\begin{equation}
\frac{d\rho}{dt}=
-\frac{i}{\hbar}\left[H,\rho \right]
- \int d^2r_1 \int d^2r_2\; K({\bf r}_1-{\bf r}_2)\;
[n({\bf r}_1),[n({\bf r}_2),\rho]] , 
\label{master2D}
\end{equation}
where $H$ is the self-Hamiltonian of the system of $N$ atoms in a 2D
trap,
\begin{equation}
H=\int d^2r\; \left(- \frac{\hbar^2}{2m} \nabla \Psi^{\dagger} \nabla \Psi+
U({\bf r}) \Psi^{\dagger} \Psi + \frac{g_{2D}}{2} \Psi^{\dagger}
\Psi^{\dagger}
\Psi \Psi \right)\;.
\end{equation}
A nonlinear stochastic (It\^{o}) unraveling of the master equation
(\ref{master2D}) is
\begin{eqnarray}
d |\Psi\rangle &=& -\frac{i}{\hbar} dt H |\Psi\rangle 
- dt \int d^2r_1 \int d^2r_2 \;  K({\bf r}_1-{\bf r}_2)\;
\Delta n({\bf r}_1)\;  \Delta n({\bf r}_2)\; 
|\Psi\rangle
+ \int d^2r\; dW({\bf r})\; \Delta n({\bf r})\;  |\Psi\rangle \;.
\label{un}
\end{eqnarray}
Here $\Delta n({\bf r})= n({\bf r})-\langle\Psi|n({\bf r}) |\Psi\rangle$
and the Gaussian noises have correlators $\overline{dW({\bf r})}=0$ and
$\overline{ dW({\bf r}_1) dW({\bf r}_2) } = 2 dt K({\bf r}_1-{\bf r}_2)$.
This unravelling corresponds to phase contrast measurement of the density of the condensate. The
evolution of $|\Psi\rangle$ given by Eq.(\ref{un}) describes a single realization of the experiment
\cite{Carmichael,Wiseman,Gatarek,Plenio}.

For a single atom, $N=1$, described by a wavefunction 
$\phi(t,{\bf r}) = \langle {\bf r} | \Psi \rangle$, 
the stochastic Schr\"odinger equation takes the form
\begin{equation}
d\phi({\bf r}) \;=\;
- \frac{i}{\hbar} dt\;
\left[\;
-\frac{\hbar^2}{2m}\nabla^2\;+\;
U({\bf r})
\right]\;
\phi({\bf r})\;+\;
\left[
dW({\bf r})-
\int d^2r'\;
|\phi({\bf r}')|^2\;
dW({\bf r}')\;
\right]\;
\phi({\bf r})\;+\;{\rm C.T.}\;.
\label{SGPEN1cont}
\end{equation}
The second term in brackets follows from the last term in Eq.(\ref{un}), 
while C.T. (counterterm)  comes from the second term in (\ref{un}). 
The latter is necessary to conserve the norm 
$\int d^2r\;|\phi({\bf r})|^2=1$, and it is given by 
\begin{equation}
{\rm C.T.}\;=\;
dt\;
\left[\;
-\;K(0)\;+\;
2\;\int d^2r_1 \;
K({\bf r}-{\bf r}_1)\;
|\phi({\bf r}_1)|^2\;-\;
\int d^2r_1 \int d^2r_2\;
|\phi({\bf r}_1)|^2\;
K({\bf r}_1-{\bf r}_2)\;
|\phi({\bf r}_2)|^2\;
\right] \phi({\bf r}).
\label{counterterm}
\end{equation}
This counterterm is badly nonlocal. Fortunately, to implement the
stochastic terms numerically one can use

\begin{eqnarray}
d\;|\phi({\bf r})|^2 &=&
2\;dt\;
\left[
dW({\bf r})-
\int d^2r'\;
|\phi({\bf r}')|^2\;
dW({\bf r}')\;
\right]\;
|\phi({\bf r})|^2 \nonumber \\
&& +
\overline{
\left[
dW({\bf r})-
\int d^2r'\;
|\phi({\bf r}')|^2\;
dW({\bf r}')\;
\right]^2 }\;
|\phi({\bf r})|^2+ 
\left[ \phi^{\star}({\bf r}) ({\rm C.T.}) + {\rm c.c.} \right]
\nonumber\\
&& =
2\;dt\;
\left[
dW({\bf r})-
\int d^2r'\;
|\phi({\bf r}')|^2\;
dW({\bf r}')\;
\right]\;
|\phi({\bf r})|^2 ,
\label{dp}
\end{eqnarray}
where the C.T. is used to cancel out the average of the stochastic terms
squared. The stochastic terms affect the modulus of $\phi({\bf r})$ only,
so Eq.(\ref{dp}) is all that one needs to implement them. Eq.(\ref{dp})
manifestly conserves the norm.

  For a weak measurement, when the squeezing of the quantum state is small, 
one can assume that the stochastic conditional state of N atoms in
Eq.(\ref{un}) is a product mean-field state
\begin{equation}
|\Psi\rangle = \frac{1}{\sqrt{N!}} 
\left[
\int d^2r\; \phi(t,{\bf r}) \Psi^{\dagger}({\bf r})
\right]^N |0\rangle ,
\end{equation}
with all the $N$ atoms in the condensate wavefunction $\phi(t,{\bf r})$. 
The latter evolves according to an It\^{o} stochastic
Gross-Pitaevskii (SGP) equation 
\begin{eqnarray}
d\phi({\bf r}) &=&
-\frac{i}{\hbar} dt\;
\left[\;
-\frac{\hbar^2}{2m}\nabla^2\;+\;
U({\bf r})\;+\;
(N-1) \;g_{2D}\;
|\phi({\bf r})|^2\;
\right]\;
\phi({\bf r}) \nonumber \\
&& +
\left[
dW({\bf r})-
\int d^2r'\;
|\phi({\bf r}')|^2\;
dW({\bf r}')\;
\right]\;
\phi({\bf r})\;+\;{\rm C.T.}\;.
\label{SGPEcont}
\end{eqnarray}
In comparison to the case of $N=1$ (see Eq.(\ref{SGPEN1cont})), this SGP
equation
has the usual extra nonlinear term $(N-1) g_{2D}|\phi({\bf r})|^2$ but the
stochastic terms containing $dW$ are the same as for a single atom
described by Eq.(\ref{SGPEN1cont}). This may seem strange because a state
of $N\gg 1$ atoms in the same quantum state can be found much faster than
a state of a single atom. One might expect the stochastic terms, which
describe backaction of the measurement on the state of the system, to
become stronger with increasing $N$. The backaction on the $N$-atom state
is indeed stronger than on the single atom state but its effect is divided
over $N> 1$ atoms instead of one. The backaction effect on the mean-field
state $\phi({\bf r})$ of each of the $N$ atoms is the same as the
backaction on the quantum state of a single atom. Thus the stochastic
terms for $N>1$ are the same as for $N=1$.

As a final remark, we mention that a stochastic Gross-Pitaevskii equation 
has been proposed and studied for a quite different goal, namely 
to describe single trajectories through unraveling of the exact 
N-body quantum evolution of a Boson system \cite{Carusotto}. In the latter 
case the interpretation of the underlying  stochasticity is obtained in 
terms of the randomness attributable to each quantum trajectory, to be 
confronted with the stochasticity that in our case is instead due to the 
opening of the condensate to a particular environment, namely the measurement 
apparatus.


\section{SGP versus exact quantum evolution: measurements in a double well}

In this Section we want to test the validity of the SGP equation in a
significant but simple situation. 
To this end we consider the double well
problem in the two mode approximation, and compare the quantum dynamics
with measurement backaction with the dynamics given by the SGP equation. The
Hamiltonian of the model is

\begin{equation}
H = \epsilon (a_1^{\dagger} a_1 + a_2^{\dagger} a_2) - 
\hbar \omega (a_1^{\dagger} a_2 + a_2^{\dagger} a_1) +
\frac{G}{2} [ (a_1^{\dagger})^2 a_1^2 + (a_2^{\dagger})^2 a_2^2 ] ,
\end{equation}
where $\epsilon$ is the mode frequency (assumed to be the same for both
modes), $\omega$ is the tunneling angular frequency, 
and $G$ is the two-particle
interaction strength. We perform phase-contrast imaging on each site, and
we assume that the kernel resolution is much shorter than the distance
between sites, so that there is no cross term due to the measurement. The
It\^{o} version of the stochastic Schr\"odinger equation for the state ket
reads

\begin{equation}
d |\Psi_{\rm Q} \rangle = dt \left[ -\frac{i}{\hbar} H - 
\frac{S}{2} (n_1 - \langle n_1 \rangle)^2 -
\frac{S}{2} (n_2 - \langle n_2 \rangle)^2 \right] 
|\Psi_{\rm Q} \rangle
+  dW_1 (n_1 - \langle n_1 \rangle) |\Psi_{\rm Q}\rangle
+  dW_2 (n_2 - \langle n_2 \rangle) |\Psi_{\rm Q}\rangle ,
\end{equation}
where the noise satisfies $\overline{dW_{\alpha}}=0$ and
$\overline{dW_{\alpha} dW_{\beta}}= 2  K_{\alpha\beta} dt$, with
$\alpha,\beta=1,2$, and $K_{\alpha\beta}=S \delta_{\alpha\beta}$ 
the measurement kernel. By assuming a total of $N$ atoms, distributed between 
the two sites, $N=N_1+N_2$, we can expand the state in terms of Fock 
states at each well 

\begin{equation}
| \Psi_{\rm Q} \rangle = \sum_{k=0}^N \psi_k(t) |k, N-k \rangle .
\end{equation}

We solve numerically the corresponding
equation for the coefficients $\psi_k(t)$ starting from a mean
field state with equal mean populations in each well,
$|\Psi\rangle_{t=0}=\frac{1}{\sqrt{N!}}(\frac{1}{\sqrt{2}} a_1^{\dagger} +
\frac{1}{\sqrt{2}} a_2^{\dagger})^N |0\rangle $. The equation provides us
with the full quantum evolution including the measurement backaction.  
We want to
compare it with the one that results from the stochastic GP equation. The
GP state is

\begin{equation}
| \Psi_{\rm GP} \rangle = \frac{1}{\sqrt{N!}} 
[ \phi_1(t) a_1^{\dagger} + \phi_2(t) a_2^{\dagger}]^N |0 \rangle ,
\end{equation}
where the wavefunctions $\phi_1$ and $\phi_2$ satisfy the following set of 
It\^{o} SGP equations,  that follow from 
Eqs.(\ref{counterterm},\ref{SGPEcont})
\begin{eqnarray}
d \phi_1 &=& -\frac{i}{\hbar} dt [(\epsilon + (N-1) G |\phi_1|^2) 
\phi_1 - \omega \phi_2] 
	     + dW_1 (1-|\phi_1|^2) 
	     + S dt [ -1+2 |\phi_1|^2 -(|\phi_1|^4+|\phi_2|^4) ] , 
\nonumber \\
d \phi_2 &=& -\frac{i}{\hbar} dt [(\epsilon + (N-1) G |\phi_2|^2) 
\phi_2 - \omega \phi_1] 
	     + dW_2 (1-|\phi_2|^2) 
	     + S dt [ -1+2 |\phi_2|^2 -(|\phi_1|^4+|\phi_2|^4) ] ,
\end{eqnarray}
and we take the same initial state as in the quantum evolution, namely a
coherent state $\phi_1(0)=\phi_2(0)=1/\sqrt{2}$ with balanced populations
$n_1(0)=n_2(0)$. In the Fig.1 we compare the time evolution of the
quantum and SGP mean populations in one well for increasing values of the
measurement strength.  In these simulations the total number of particles
was $N=100$, and the nonlinear coupling was $G=0$. In this case, the
Hamiltonian involves only one-body terms, so the mean field evolution
(based on coherent states) must exactly coincide with the quantum one for
the case of zero measurement ($S=0$). For small measurement strengths, $n
S/\omega \ll 1$ (here $n$ denotes the average number of particles per
site), the quantum state of the condensate is still, to a high degree of
accuracy, a coherent state, so the SGP evolution and the quantum one
coincide. In Fig. 1a we see that the agreement is good even for times much
larger than $1/n S$.  This timescale is relevant for the strong
measurement case of the previous Section, since it sets the time after which
an asymptotic number squeezed state is reached.  As we increase the
measurement strength and reach $n S/\omega \geq 1$ (see Fig. 1b), the mean
number of particles per well given by the SGP and the quantum evolution
depart appreciably, not surprisingly since such strong measurements
squeeze the quantum state of the condensate driving it outside the
description in terms of coherent states, {\it i.e.} the associated basis
for the Gross-Pitaevskii equation.  

\begin{figure}[h]
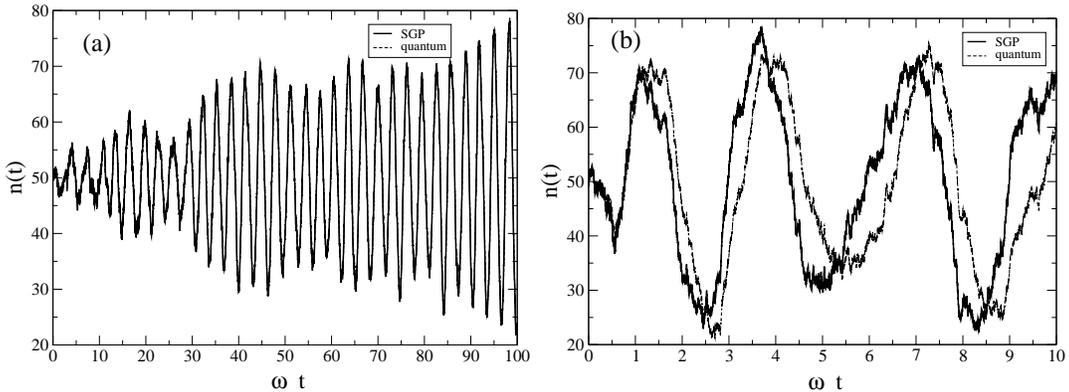
 
\centering
\epsfig{file=fig1a.eps, width=7cm, clip=} 
\epsfig{file=fig1b.eps, width=7cm, clip=} 
\caption{Comparison between the SGP and the exact quantum evolution for
weak measurements on a non-interacting Bose condensate. The population
$n(t)$ in a given well is plotted as a function of time for a single
stochastic realization, expressed in units of cycles of the Rabi
oscillations. The total number of particles is $N=100$. 
(a) The SGP evolution and the quantum one 
coincide for an effective measurement strength $nS/\omega=10^{-1}$ even 
for times much greater than $1/nS=10/\omega$. (b) The departure of the
SGP dynamics from the exact one at the time $1/nS=1/\omega$ is evident 
for a stronger measurement coupling $n S/\omega=1$.}
\label{figure1} 
\end{figure}

The Gross-Pitaevskii evolution can depart from the quantum one not only
due to the measurement backaction but also due to the nonlinearity of the
interactions. In Fig. 2a we show the SGP and quantum evolutions for
$S/\omega=10^{-3}$ and $G/\omega=10^{-3}$, corresponding to the same
initial state as in previous figures. We see that the inclusion of the
nonlinearity causes the SGP evolution to depart from the quantum one. To
gain further insight we introduce a quantity which measures the
depleted fraction of atoms from the {\it best} mean field state, i.e. a
mean field state that is the closest to the exact quantum state. Its
definition is

\begin{equation}
D = \stackrel{{\rm min}}{\{A,\phi\}} \left(1 - \frac{1}{N} 
\langle \Psi_Q | c^{\dagger} c | \Psi_Q \rangle \right)
\end{equation}
where the operator $c$ is $c=\sqrt{A} a_1 + e^{i \phi} \sqrt{1-A^2} a_2$.
In Fig.2b we plot this depletion for the same simulation of Fig. 2a. The
depletion in Fig.2b is small, less than one atom is depleted from the
condensate, but, as we see in Fig.2a, the SGP evolution departs from the
exact evolution. This departure is attributable to the inclusion of the
nonzero $G$. 

\begin{figure}[h]
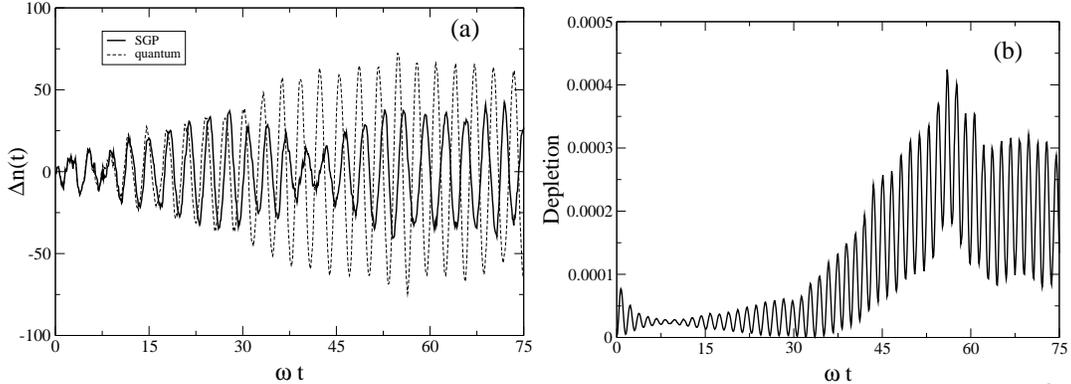

\centering
\epsfig{file=fig2a.eps, width=7cm, clip=}
\epsfig{file=fig2b.eps, width=7cm, clip=}
\caption{Comparison between the SGP and the exact quantum evolution for a
weakly interacting ($G/\omega=10^{-3}$) Bose condensate subjected to a
weak measurement $n S/\omega=10^{-1}$. (a) The population difference
$\Delta n(t)=n_2(t)-n_1(t)$ is plotted for a single stochastic
realization. Initially the condensate is in a coherent state with equal
populations in each mode, $n_1(0)=n_2(0)=50$. 
(b) Depletion from a mean field state for the same realization as in (a).}
\end{figure}

In Fig.3a we show the GP and quantum evolution for $S=0$ and
$G/\omega=10^{-3}$. Since no measurements are performed, an initial
balanced population ($n_1(0)=n_2(0)$) would remain balanced for all times,
both at the GP and quantum level. For this reason we take an initial
unbalanced population, $n_1(0)/n_2(0)=3/2$, which due to the hopping term
triggers Rabi oscillations between the two wells (just as the measurement
did when we took initial balanced populations). It follows from the figure
that the GP dynamics departs from the quantum one. We plot in Fig. 3b the
depletion corresponding to Fig. 3a. Again, this depletion remains small,
less than one atom is depleted from the condensate.

\begin{figure}
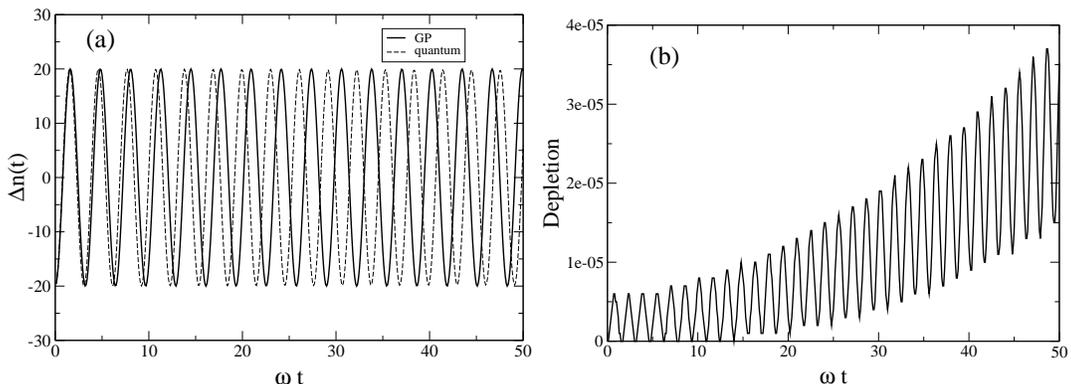

\centering
\epsfig{file=fig3a.eps, width=7cm, clip=}
\epsfig{file=fig3b.eps, width=7cm, clip=}
\caption{Comparison between the standard GP and the exact quantum
evolution for a weakly interacting ($G/\omega=10^{-3}$) unmeasured
Bose condensate.
(a) The population difference $\Delta n(t)=n_2(t)-n_1(t)$ is plotted 
as a function of time. Initially the condensate is in a coherent
state with unbalanced populations, $n_1(0)=60$, $n_2(0)=40$. 
(b) Depletion from a mean field state for the same realization as in (a).}
\end{figure}

 As we can see in Figs.2,3, both  the SGP and GP evolutions depart from the
exact evolution even in the weak measurement and interaction limits when
the depletion from the condensate is small. The derivation of the
backaction terms in the SGP equation requires only one assumption: all the
atoms are in the condensate. In contrast, the derivation of the
interaction terms (both for SGP and GP equations) not only assumes that
all the atoms are in the condensate but also makes further approximations
to describe the evolution of the condensate wavefunction $\phi$. This is why
$G \neq 0$ causes both the SGP and GP evolutions to depart from the exact
quantum evolution even for negligible depletion. However, even after the
departure the SGP and GP evolutions remain qualitatively similar to the
exact evolution, see Figs. 2a and 3a.

 To summarize, the accuracy of the SGP equation is limited by the weak
measurement condition, $nS/\omega<1$, and by the nonlinear interaction.
For a weak measurement the accuracy of the SGP equation is the same as
that of the GP equation. In the next Section we apply the SGP equation to
show how the measurement can trigger the unwinding of a dark soliton
\cite{DS,DS1} in a realistic experimental setup.


\section{Measurement-induced unwinding of a dark soliton }

  Once delimited the validity of the stochastic Gross-Pitaevskii equation 
in the simple situation of a two-mode system we can apply it to the more 
complex case of imaging of a condensate state with a nontrivial phase 
such as a soliton. Even in the limit of weak
measurement, with the solution still approximable in terms of
Gross-Pitaevskii coherent states, the effect of the measurement is present
and affects the observable conjugate to the atom number, i.e. the phase of
the condensate.  As an example of application of the SGPE let us consider
an isotropic harmonic 2D trapping potential $V({\bf r})=(m \Omega^2/2)
(x^2+y^2)$.  The condensate is assumed to be in the Thomas-Fermi limit of
strong repulsive interaction where the ground state wavefunction can be
well approximated by

\begin{equation}
\phi_{GS}({\bf r})\;=\;
\sqrt{\frac{\mu -V({\bf r})}{Ng_{2D}}}\;.
\label{GSTF}
\end{equation}
The constant $\mu=\Omega\sqrt{N g_{2D} m/\pi}$, the chemical potential, is
chosen so that the wavefunction is normalized to $1$. Let us use as the 
initial state a dark soliton \cite{DS} imprinted on the Thomas-Fermi
ground state

\begin{equation} 
\phi(t=0,{\bf r})\;=\; 
\tanh\left( \frac{x}{l} \right)\;
\sqrt{\frac{\mu'-V({\bf r})}{Ng_{2D}}}\;. 
\label{psi0}
\end{equation} 
Here $l=0.6 \mu{\rm m}$ is the healing length at the peak density in the
ground state (\ref{GSTF}).

  In our numerical simulations we assume the following parameters relevant
for $^{87}$Rb: mass $m=1.4\times 10^{-25}$kg, scattering length $a=5.8$nm,
$\chi_0=10^{-23}$m$^3$, wavelength $\lambda=780$nm. The width of the
gaussian $\Lambda(z)$ is assumed to be $\xi=10\mu$m. With these parameters
the resolution of the kernel is $\Delta r=5\mu$m. We assume a 2D harmonic
trap frequency $\Omega=2\pi\times 10 {\rm s}^{-1}$. With $N=5\times 10^5$
we get a 2D Thomas-Fermi radius of $31\mu\rm{m}$. We also assume the laser
intensity of $I=10^{-4} {\rm mW}/{\rm cm}^2$.

  The former parameters give a weak measurement strength in the sense
discussed in Sections III and IV, so that the use of the SGP equation is
justified.  Indeed, this continuous problem can be mapped on the lattice
model Eq.(\ref{lattice}), where the lattice constant is the kernel
resolution $\Delta r$. Given the Thomas-Fermi radius of $31\mu\rm{m}$, and
$\Delta r=5\mu$m, we estimate the lattice to be composed of $100$ sites
with an average of $n=5000$ atoms per site. The measurement strength is
equal to $S=7\times 10^{-5} {\rm s}^{-1}$, and the effective hopping
frequency for the lattice model is $\omega=14 {\rm s}^{-1}$. Therefore $n
S/\omega \approx 2\times 10^{-2} \ll 1$, thus confirming a weak
measurement regime.

  To simulate the continuum SGP equation (\ref{SGPEcont}) we discretized
it using a lattice constant $2 \pi$ times smaller than the kernel
resolution $\Delta r$. The program uses a split step method: the fast
Fourier transform was used to carry out the time integration of the
kinetic term and of the nonlocal terms involving the kernel, and the
potential and nonlinear coupling terms were integrated in time in the
position representation. A cross-section along the $x$-axis through the
probability density of the initial state (\ref{psi0}) is shown in (a) of
Fig.4. Cross sections through probability densities at later times after
the probe light beam has been sent on the condensate are shown in cases
(b) and (c) of Fig.4.  For comparison, the time evolution without the
measurement does not result in any soliton unwinding.

\begin{figure}[h]
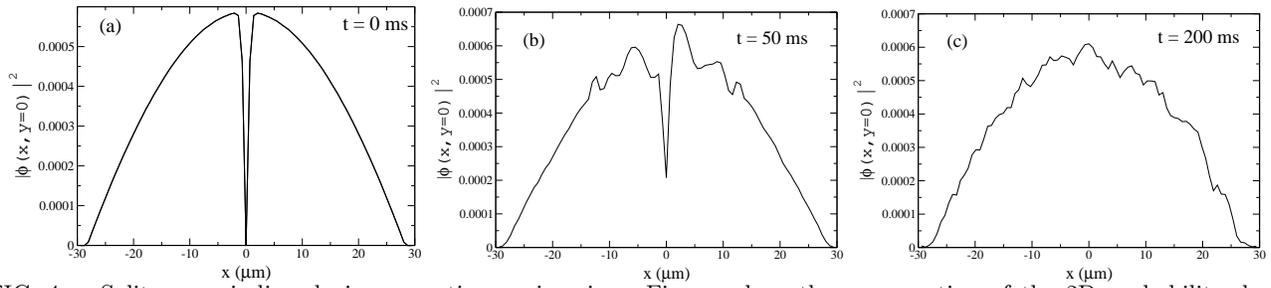
 
\centering 
\epsfig{file=fig4a.eps, width=5.5cm, clip=}
\epsfig{file=fig4b.eps, width=5.5cm, clip=}
\epsfig{file=fig4c.eps, width=5.5cm, clip=}
\caption{\label{stage0} Soliton unwinding during a continuous imaging.  
Figures show the cross section of the 2D probability density
$|\phi(x,y)|^2$ of the Bose condensate along the $x$-axis at $t=0$ ms (a),
$t=50$ ms (b), and $t=200$ ms (c) since the beginning of the continuous
imaging.  The dark soliton is greying progressively and in the third
profile there is no trace of the initially imprinted pattern.  
The laser intensity is $I=10^{-4} {\rm mW}/{\rm cm}^2$, 
and the condensate wavefunction is normalized to unity in 2D.}
\end{figure} 

  The soliton unwinds after roughly $50\;$ms. This time is much shorter
than the depletion time $t_2=10^4\;$s discussed at the end of Section II.  
The measurement induces soliton unwinding much earlier that any detectable
depletion of atoms occurs. Figure \ref{stage0} suggests that the
unwinding will manifest itself by filling up the soliton core with atoms.
Such a greying of the dark soliton can also occur through a different
mechanism that involves collisions between condensate atoms. In
Ref.\cite{DKS} it was demonstrated that the dark soliton can grey on a
timescale of tens of ms because its core fills up with non-condensed atoms
(quantum) depleted from the condensate as a result of atomic collisions
between condensed atoms. This result is supported by Ref.\cite{Law} where
it was shown that the quantum state with minimal depletion has depletion
strongly concentrated in the soliton core.  This quantum depletion process
does not unwind the soliton: the condensate remains in the soliton state
with a phase jump of $\pi$. The phase jump or its unwinding could be
detected by interference between two condensates, one of them in a ground
state and the other with a soliton \cite{INT}. 
A simpler way to verify that observed
greying is due to the measurement induced unwinding is to change the
measurement strength in a certain range and see if the greying time
depends on the imaging laser intensity. We simulated the soliton unwinding
for a range of measurement strengths. Fig. 5 shows the unwinding time
$\tau$ versus the laser intensity $I$. The time $\tau$ is defined as that
for which the density at $x=0$ in Fig. 4 achieves $10\%$ of the maximal
density.

\begin{figure}[h]
\centering
\epsfig{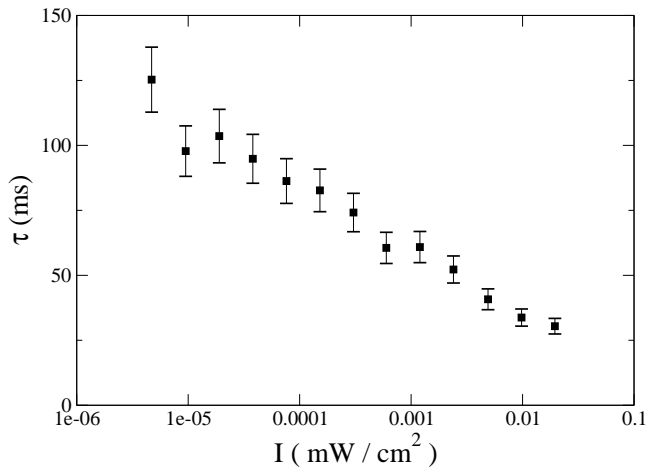}
\caption{ The unwinding time $\tau$ as a function of the
laser intensity $I$.}
\end{figure}


\section{Conclusions}

Quantum measurement theory has been applied to the dispersive imaging 
of a Bose-Einstein condensate. In the strong measurement limit the condensate 
is irreversibly driven into non-classical states with reduced number 
fluctuations. In the opposite limit of weak measurement the condensate 
can be approximately described for short timescales through a stochastic 
counterpart of the Gross-Pitaevskii equation. The latter has been applied 
for the study of the dynamics induced by dispersive imaging on a condensate 
prepared in a soliton state. The proposed model, besides allowing to 
intentionally design selective manipulation of the condensate state for 
instance to quench vortices without introducin appreciable depletion, 
could also lead to a better understanding of quantum phase transitions 
\cite{Fisher,Sachdev} in Bose condensates \cite{MOTTBEC}.


\section{Acknowledgements} 

We would like to thank Ivan Deutsch for calling our attention to Refs \cite{Kuzmich,Kuzmich1}.
The work of D.D. and J.D. was supported in part by NSA.


\end{document}